\documentclass[10pt,a4paper,superscriptaddress,showpacs,showkeys,preprintnumbers]{revtex4}
\usepackage{graphicx}
\usepackage{subfigure}
\begin{document}

\title{Experimental Study of Pressure Influence on Tunnel Transport into 2DEG}

\author{E.M. DIZHUR}
\email[]{Phone: 7(095)3340590, Fax: 7(095)3340012,
e-mail: dizhur@east.ru, dizhur@ns.hppi.troitsk.ru}
\author{A.N. VORONOVSKY}
\affiliation{Institute for High Pressure Physics of the RAS, Troitsk 142190,
Moscow Reg., Russia}
\author{I.N. KOTEL'NIKOV}
\author{S.E. DIZHUR}
\email[]{e-mail: sdizhur@cplire.ru}
\affiliation{Institute of Radioengineering and Electronics of the RAS,
Moscow 101999,Mokhovaya St. 11, Russia}
\author{M. N. FEIGINOV}
\affiliation{Institute of Radioengineering and Electronics of the RAS,
Moscow 101999,Mokhovaya St. 11, Russia}
\affiliation{Technische Universitaet Darmstadt, Institut fuer
Hochfrequenztechnik, Merckstr.25, D-64283 Darmstadt, Germany}

\begin{abstract}
We present the concept and the results of pilot measurements of tunneling in
a system \mbox{Al/$\delta_{Si}$-GaAs} under pressure up to $2 \, GPa$ at
$4.2 \, K$.
The obtained results may indicate the following: the barrier height for
\mbox{Al/$\delta$-GaAs}
equals
to $0.86 \, eV$ at $P=0$ and its pressure coefficient is $3\,  meV/kbar$;
charged impurity density in the delta-layer starts to drop from
$4.5\times 10^{12} \, cm^{-2}$ down to $3.8\times 10^{12} cm^{-2}$ at
about $1.5 \, GPa$; metal-insulator transition may occur in 2DEG at
about $2 \, GPa$.
\end{abstract}

\pacs{62.50.+p; 73.21.-b; 73.30.+y; 73.40.Gk}

\keywords{pressure; $\delta$-dopped layer; tunnelling}

\maketitle

\section{Introduction}
In the very vicinity of metal-insulator transition charge
transport properties of a 2DEG depend not only on the interplay between the
electron density and disorder but also on the influence of quasi-particle
interaction.
Tunneling measurements under pressure may provide a useful
experimental approach to this fundamental problem by the following reasons.
Contrary to magnetotransport measurements that obviously deal with the
carriers only in filled subbands of spatial quantization, tunneling allows
also to get an information on energy spectra of empty higher lying subbands.
Many-particle effects (exchange and/or electron-phonon interactions)
manifest themselves in tunnel characteristics either through the changes
in a barrier shape \cite{1} or as self-energy effects \cite{2}.
Examples are the specific singularities at biases near ($36.5 \, mV$,
corresponding to LO phonons in GaAs and the so called zero bias anomaly
originated from the interaction between the electrons revealing as a peak
of tunnel resistance \cite{1, 3}).
High pressure may be used as a tool that allows to change substantially the
electron density as well as the energy levels in a quantum well formed by
selective doping of a narrow layer near the \mbox{Me/GaAs} interface,
maintaining concentration and distribution of static scattering centers
(impurities, dislocations etc) practically constant.
An additional mechanism of the pressure influence on the charged impurity
density may be related to the existence of DX centers as it was shown
earlier in magnetotransport measurements \cite{4}.
In this paper we present a first attempt (to our knowledge) to measure
normal tunneling in a system \mbox{Al/$\delta_{Si}$-GaAs} with 2DEG formed
by the delta-doped layer close to the interface under pressure
up to $2 \, GPa$ at $4.2 \, K$.

\section{Experimental}
\begin{figure}[h]
\centering
\subfigure[][]{\label{1-a}\includegraphics[width=6cm]{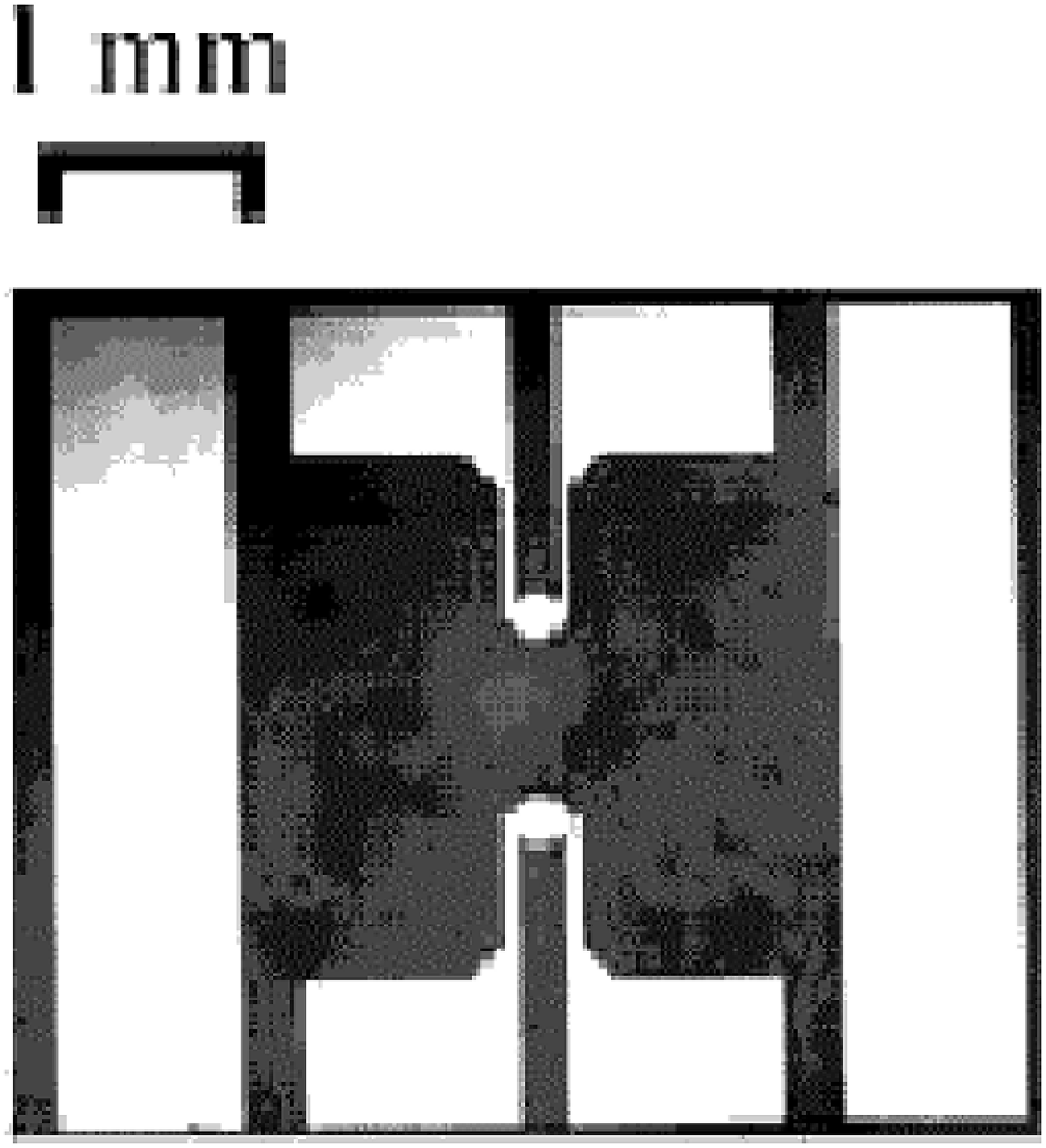}}
\hspace{3cm}
\subfigure[][]{\label{1-b}\includegraphics[width=6cm]{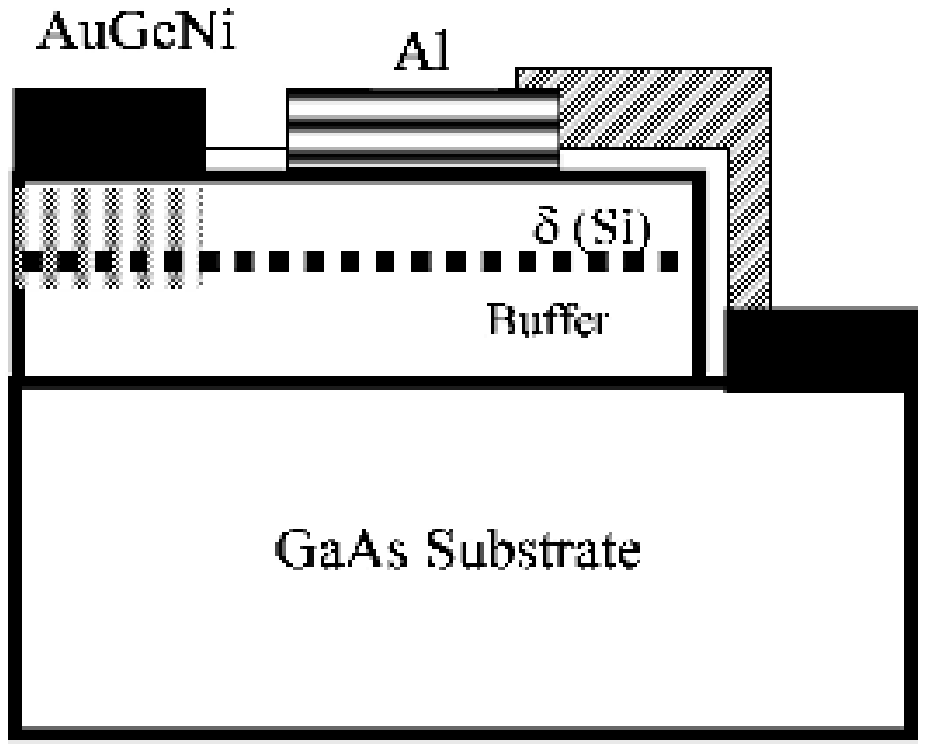}}
\caption{Plain view (a) and schematic cross-section (b) of the sample Z1B8.}
\label{1}
\end{figure}

The tunnel structures, shown in FIG.~\ref{1}, were fabricated in
IRE RAS following the procedure described in \cite{3}. On (100)
GaAs substrate by the MBE method an undoped buffer layer ($p\sim
5\times 10^{15} \, cm^{-3}$) was grown. The delta-doped layer with
width about $3 \, nm$ at the depth of $20 \, nm$ from
\mbox{Al/GaAs} interface containing $5.2\times 10^{12} \, cm^{-2}$
Si atoms was formed at $570 \,^{\circ}C$. At the final stage of
growth cycle, an Al film of $80-100 \, nm$ width was deposited
from the Knudsen cell in the same MBE chamber.
Photolithographically shaped tunnel junctions were supplied with
Au-Ge-Ni alloy ohmic contacts to the delta-layer.

\begin{figure}[h]
\centering
\subfigure[][]{\label{2-a}
\includegraphics[width=7cm,keepaspectratio]{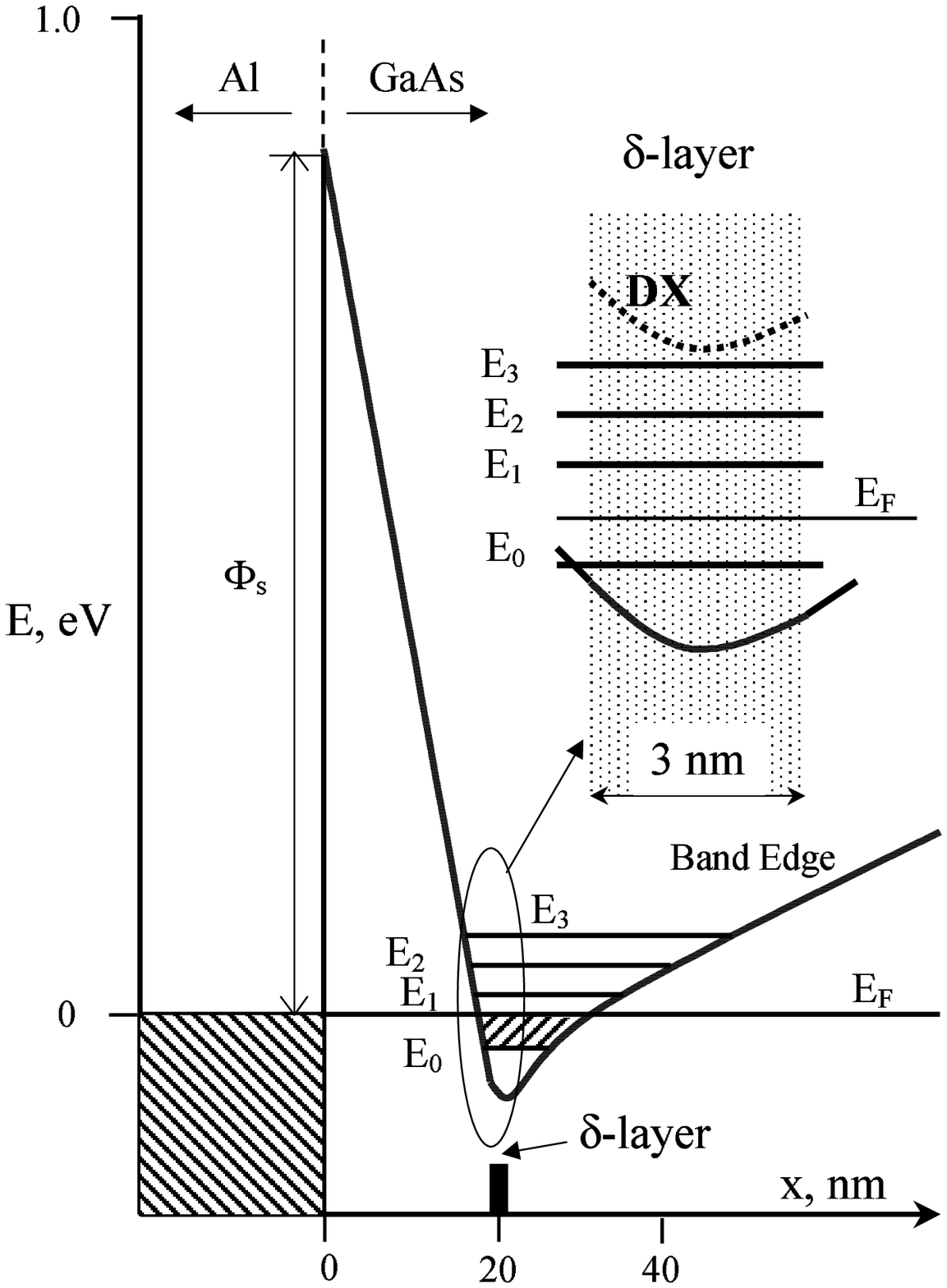}}
\hspace{2cm}
\subfigure[][]{\label{2-b}
\includegraphics[width=8cm,keepaspectratio]{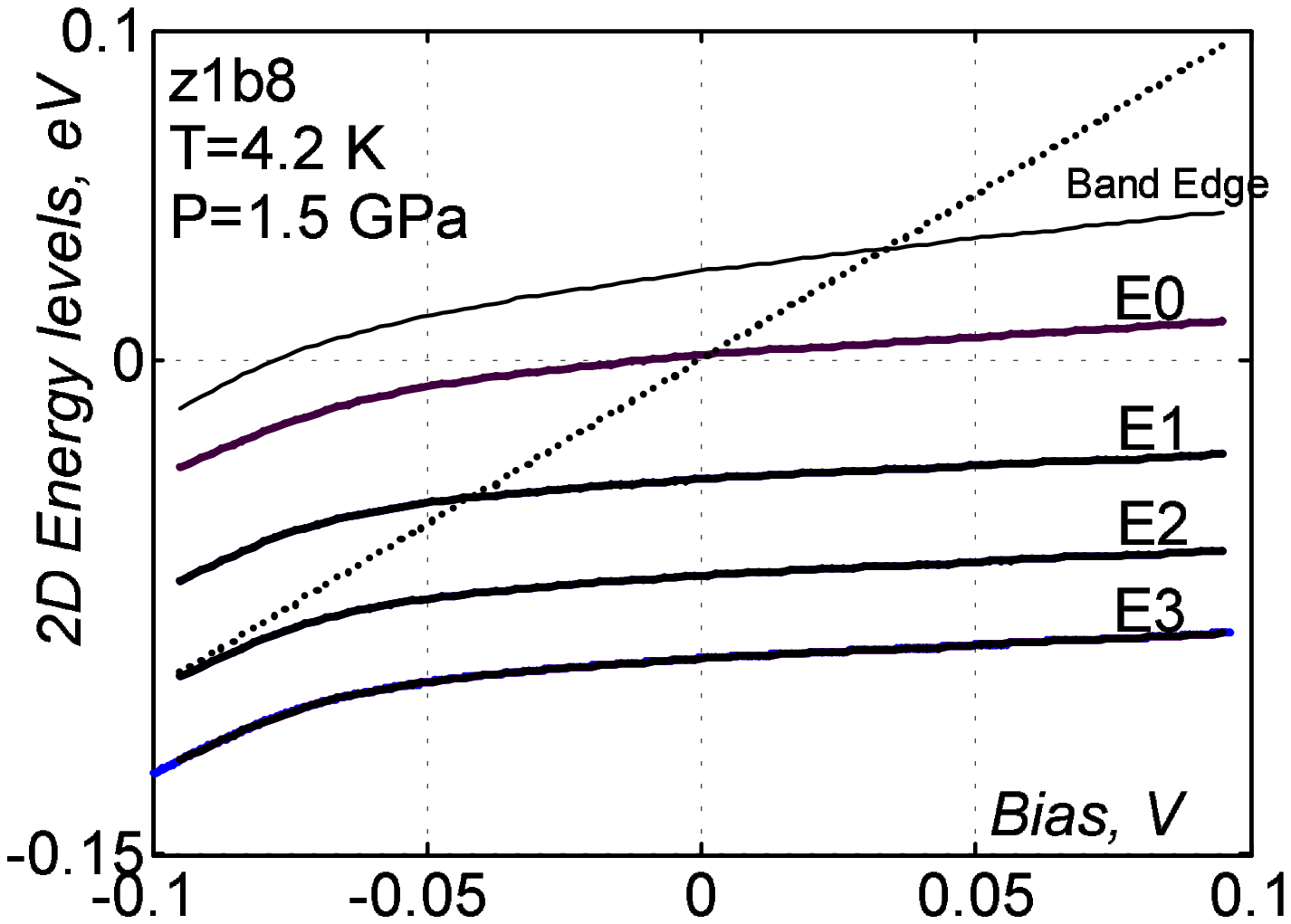}}
\caption{(a) Energy diagram in the vicinity of \mbox{Al/GaAs} interface.
Dotted line represents possible position of DX-level.\\
(b) The calculated energy of the 2D-levels relative to Fermi energy of the
metal electrode at $P=1.5 \, GPa$. The corresponding features on the tunneling
spectrum may be found graphically as the points of intersection between
the levels and the dotted line.}
\label{2}
\end{figure}

In this structure only one level is initially occupied at $T=4.2 \, K$.
The energy diagram of the sample that defines tunneling current in the
direction normal to the interface is shown on FIG.~\ref{2-a}.
The spatial distribution of the electric potential that forms both the
tunnel barrier and the quantum well depends on the surface barrier height,
on doping level in delta-plane and on volume acceptor density in the
epitaxial layer.
A system of energy levels exists in the quantum well that gives rise to
the dips on tunneling spectra corresponding to the stepwise changes in
conductance versus applied bias.
One should note that the 2D levels are changed in the biased diode as the
barrier (and QW) shape selfconsistently depends on the charge density in
the delta-layer.
FIG.~\ref{2-b} shows the variation of the 2D levels in the well with the
applied bias, calculated for $p=1.5 \, GPa$ by simultaneous solution of
Poisson and Schr\"odinger (in Hartree approximation) equations.

The pressure was generated at room temperature in a stand-alone
high-pressure cell of a piston-cylinder type filled with 40\% transformer
oil and 60\% pentane mixture as a pressure transmitting medium \cite{5}.
After
slow cooling down to low temperatures the $I(V)$ curves were measured within
the accuracy $7\frac{1}{2}$ digit DC voltmeter. The actual pressure was evaluated
by the change of the critical temperature $T_{c}$ of Sn wire placed {\it
in situ} using the expression: $\Delta T_{c} = -0.495P+0.039P^2$
(the pressure $P$ in $GPa$) \cite{6}.

\begin{figure}[h]
\includegraphics[width=10cm]{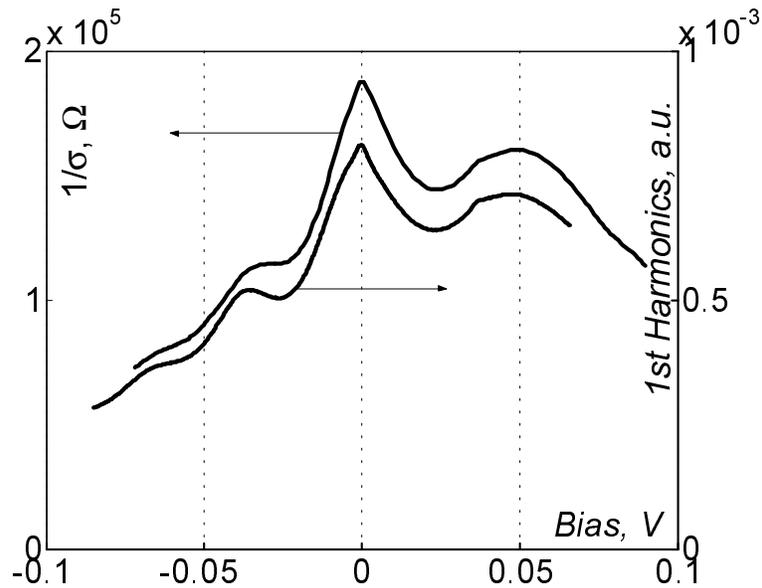}
\caption{
Comparison of DC and modulation techniques. The upper curve is the result
of numerical derivative of DC $I(V)$, the lower one is a signal of the 1st
harmonics in usual modulation technique for the same sample at zero
pressure.}
\label{3}
\end{figure}

\begin{figure}[h]
\includegraphics[height=10cm]{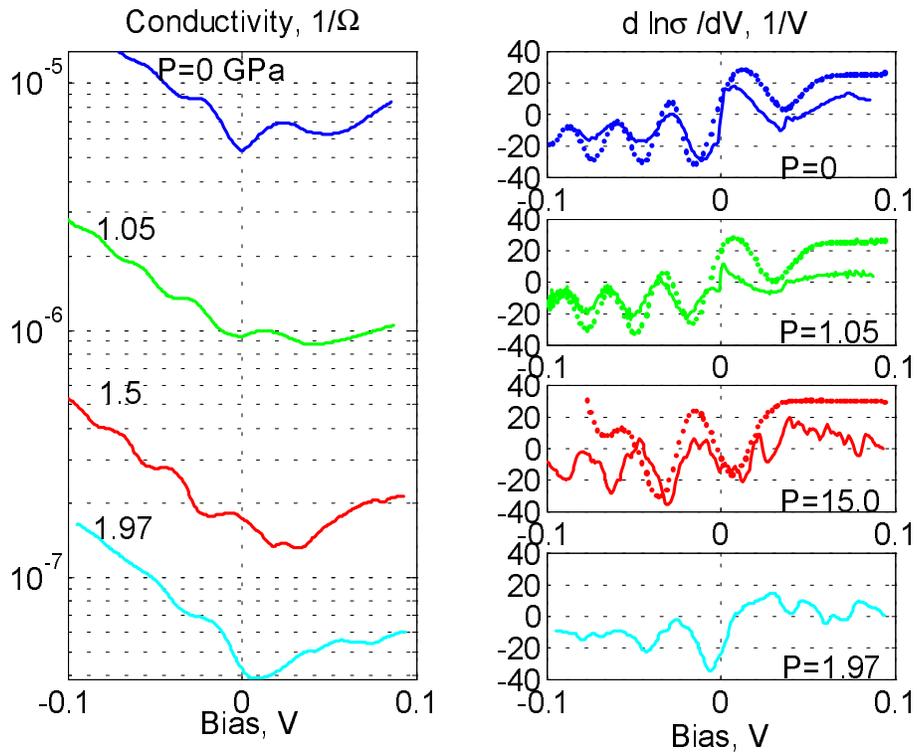}
\caption{
Conductance $\sigma$ and tunneling spectra $d\ln\sigma/dV$ at various pressures.
Dotted lines correspond to model calculations.}
\label{4}
\end{figure}

Under pressure the tunnel resistance tends to increase rapidly with
pressure \cite{7}.
This made us to confess that the usual measurement procedure based on
modulation technique is not well suited for the case of large tunnel
resistances.
Instead, we used direct current $I(V)$ measurements and subsequent
numerical derivation to reveal the fine features due to the subband
structure and many-particle effects.
The comparison of the both measurement techniques for the case of zero
pressure when the tunnel resistance is of moderate value, confirms the
validity of DC implementation, that becomes even more effective under
pressure when the resistance grows by several orders of magnitude
(FIG.~\ref{3}).
Results of $I(V)$ measurements in terms of differential tunneling
conductance as well as its logarithmic derivative, {\it i.e.} tunneling
spectra, are presented in FIG.~\ref{4}.

\section{Analysis}
Model calculations were used to evaluate the tunneling barrier shape and
the position of energy levels in the QW, and hence the tunneling spectra
and the value of tunnel resistance at zero bias $R_{0}(P)$, the latter being
a simple and convenient scalar quantity to compare the model assumptions
with reality.

From our previous studies of Shottky barrier \mbox{Me/$n^{+}$-GaAs} \cite{7}
we knew that barrier height at the \mbox{Al/GaAs} interface may change with
pressure not the same way as the energy gap in the semiconductor does.
Possible presence of DX-centers could also change effective charge state of
the impurity in the delta-layer under pressure.
Thus we considered such an ambigous at high pressure quantities as free
parameters and tried possibly fit the calculated and measured spectra and
zero bias resistances at different pressures.

We found that the best fit to the experimental spectra at pressures up to
$1.5 \, GPa$ may be attained if one assumes that the barrier height grows
at the rate of $3 \, meV/kbar$, that is considerably lower than the pressure
coefficient of the band gap $10.8 \, meV/kbar$ \cite{8}.

\begin{figure}[h]
\includegraphics[width=10cm,keepaspectratio]{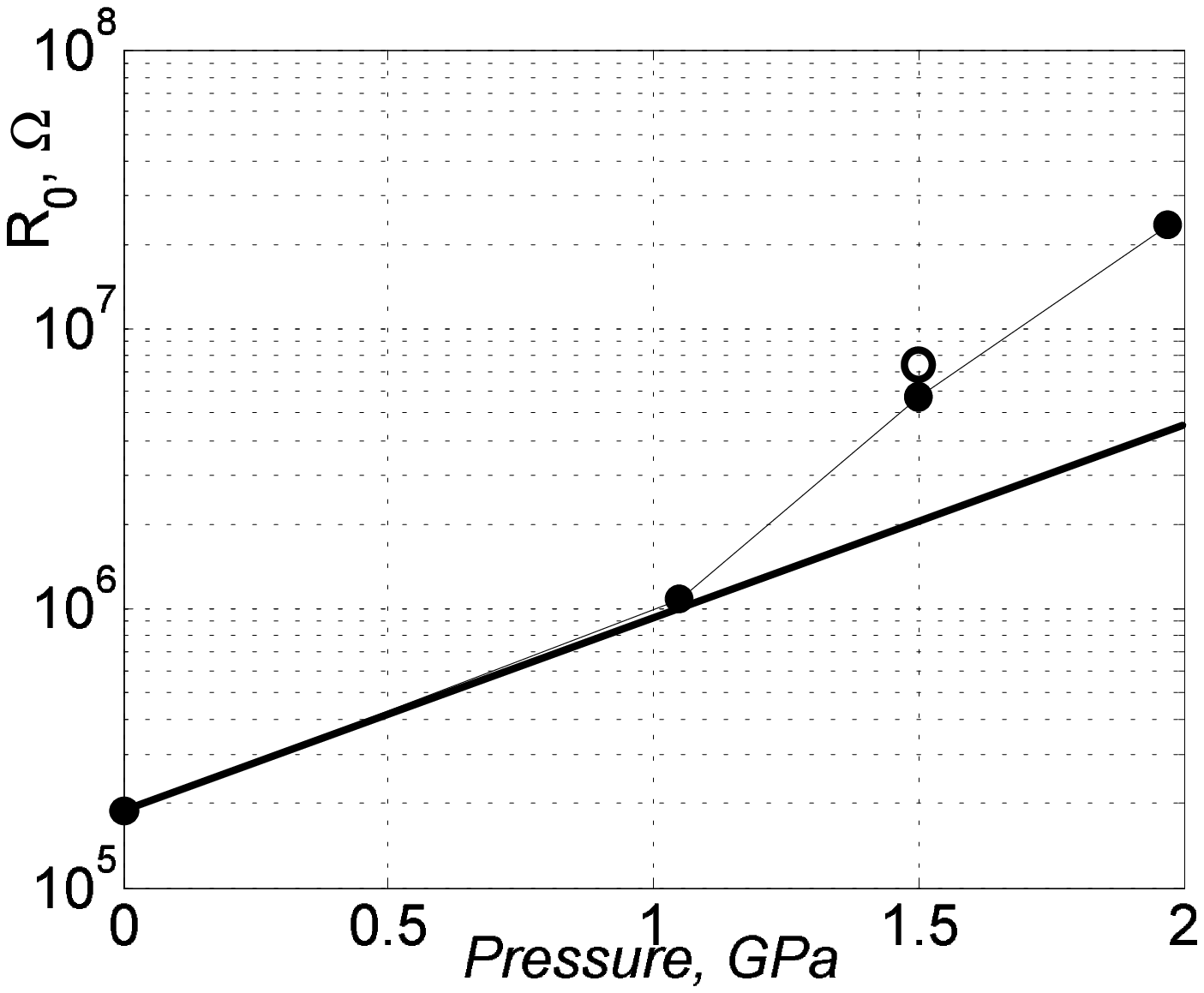}
\caption{
Tunnel resistance at zero bias. Solid symbols are experimental data, a
dashed line -- theory for constant charged impurity density in the
delta-layer, the open circle -- theory for effective charge density
$3.8\times 10^{12} \, cm^{-2}$.}
\label{5}
\end{figure}

Comparison of the zero bias resistances (FIG.~\ref{5}) shows that while a
reasonable agreement with the experiment (filled circles) may be achieved
for $1 \, GPa$ with the charged impurity density in delta-layer
$4.5\times 10^{12} \, cm^{-2}$ (same as at $P=0$), a systematic underestimate
of the calculated resistance is observed at $P=1.5\,  GPa$.
A better fit for that pressure (open circle),
that gives also closer fit of tunnel spectrum, may be obtained if one
assumes the decrease of charged impurity density in delta-layer down to
$3.8 \times 10^{12} \, cm^{-2}$ at $1.5 \, GPa$.

A possible reason for that may be related to the electron capture by
DX-centers, making them either negative \cite{9} or neutral \cite{10}, that
changes the effective charged impurity density in the delta-layer.
To become resonant with the ground state at $1 \, GPa$, DX-level at $P=0$
should be about $170 \, meV$ above the conduction band edge that is not
quite consistent with the previous data \cite{11}, obtained from
magnetotransport measurements in a deep lying delta-layer with multiple
occupied 2D-levels.

We could not obtain any reasonable model description of our experimental data
at $2 \, GPa$.
The estimation allows supposing that these measurements correspond to the
state when a metal-insulator transition occurs and the ground level is
depleted, but this cannot explain the existence of slight remnants of higher
levels in spectrum at negative biases.

\begin{acknowledgments}
The authors acknowledge financial support by Russian Foundation
for Basic Researches and the Alexander von Humboldt Foundation,
the Federal Ministry of Education and Research and the Programme
for Investment in the Future (ZIP) of the German Government.
\end{acknowledgments}

\newpage

\vspace{2cm}

\newpage

\vspace{2cm}

\newpage

\begin{thebibliography}{20}

\bibitem{1} A.~YA.~SHUL'MAN, I.~N.~KOTEL'NIKOV, N.~A.~VARVANIN, E.~N.~MIRGORODSKAYA,
\textit{Pis'ma v Zh.Eksper. Teor.Fiz.}, {\bf 73}, 573 (2001).
\bibitem{2} J.~W.~CONLEY AND G.~D.~MAHAN, \textit{Phys. Rev.}, {\bf 161}, 681 (1967).
\bibitem{3} I.~N.~KOTEL'NIKOV, V.~A.~KOKIN, YU.~V.~FEDOROV, A.~V.~HOOK, and D.~T.~TALBAEV,
\textit{Pis'ma v Zh.Eksper. Teor.Fiz.}, {\bf 71}, 564 (2000).
\bibitem{4} A.~ZRENNER, F.~KOCH, R.~L.~WILLIAMS, R.~A.~STRADLING, K.~PLOOG and G.~WEIMANN,
\textit{Semicond. Sci. Technol.}, {\bf 3}, 1203 (1988).
\bibitem{5} A.~N.~VORONOVSKY,  E.~S.~ITSKEVICH,  E.~M.~DIZHUR, and L.~M.~KASHIRSKAYA
\textit{Theses of the Conf. on "Advanced methods of pressure treatment"},
1982, Chechoslovakia, Bratislava.
\bibitem{6} L.~D.~JENNINGS, C.~A.~SWENSON,
\textit{Phys. Rev.}, {\bf 112}, 31 (1958).
\bibitem{7} E.~M.~DIZHUR, A.~YA.~SHUL'MAN, I.~N.~KOTEL'NIKOV, and A.~N.~VORONOVSKY,
\textit{Phys.Stat.Sol. B}, {\bf 223}, 129 (2001).
\bibitem{8} J.~LEYMARIE, M.~LEROUX, and G.~NEU,
\textit{Phys. Rev. B}, {\bf 42}, 1482 (1990).
\bibitem{9} D.~J.~CHADI and K.~J.~CHANG,
\textit{Phys. Rev. B}, {\bf 39}, 10063 (1989).
\bibitem{10} E.~K.~YAMAGUCHI, K.~SHIRAISHI, and T.~OHTO,
\textit{J. Phys. Soc. Japan}, {\bf 60}, 3093 (1991).
\bibitem{11} P.~M.~KOENRAAD, W.~DE~LANGE, A.~A.~P.BLOM, M.~R.~LEYS, J.~A.~A.~J.~PERENBOOM,
J.~SINGLETON and J.~H.~WOLTER,
\textit{Semicond. Sci. Technol.}, {\bf 6}, B143 (1991).
\end{thebibliography}
\end{document}